# Doping Dependence of Thermal Oxidation on n-type 4H-SiC

B.K. DAAS[*], M.M. ISLAM, I. A. CHOWDHURY, F. ZHAO, T.S. SUDARSHAN, M.V.S. CHANDRASHEKHAR

Department of Electrical Engineering, University of South Carolina,

301 S. Main St, Columbia, SC 29208

*Daas@email.sc.edu

## Abstract:

The doping dependence of dry thermal oxidation rates in n-type 4H-SiC was investigated. The oxidation was performed in the temperature range $1000^0C$ to $1200^0C$ for samples with nitrogen doping in the range of $6.5\times10^{15}/cm^3$ to $9.3\times10^{18}/cm^3$, showing a clear doping dependence. Samples with higher doping concentrations displayed higher oxidation rates. The results were interpreted using a modified Deal-Grove model. Linear and parabolic rate constants and activation energies were extracted. Increasing nitrogen led to an increase in linear rate constant pre-exponential factor from $10^{-6}$m/s to $10^{-2}$m/s and the parabolic rate constant pre-exponential factor from $10^{-9}$m$^2$/s to $10^{-6}$m$^2$/s. The increase in linear rate constant was attributed to defects from doping-induced lattice mismatch, which tend to be more reactive than bulk crystal regions. The increase in the diffusion-limited parabolic rate constant was attributed to degradation in oxide quality originating from the doping-induced lattice mismatch. This degradation was confirmed by the observation of a decrease in optical density of the grown oxide films from 1.4 to 1.24. The linear activation energy varied from 1.6eV to 2.8eV , while the parabolic activation energy varied from 2.7eV to 3.3eV, increasing with doping concentration. These increased activation energies were attributed to higher nitrogen content, leading to an increase in effective bond energy stemming from the difference in C-Si (2.82eV) and Si-N (4.26eV) binding energies. This work provides crucial information in the engineering of $SiO_2$ dielectrics for SiC MOS structures, which typically involve regions of very different doping concentrations, and suggests that thermal oxidation at high doping concentrations in SiC may be defect mediated.

## Index Terms:

4H-SiC, Doping dependence, Thermal Oxidation, Linear rate constant, Parabolic rate constant, Activation energy, Optical density

## Introduction:

4H-SiC's superior electrical properties , such as wide band gap (3.2eV), high breakdown electric field(2.2MV/cm), high saturation velocity($2.0x10^7$ cm/s), high mobility ($1000cm^2$/Vs) and high thermal conductivity (4.9W/cmK) make it a promising semiconductor material for power electronic devices in smartgrid power conversion applications [1].

Unlike other wide bandgap semiconductor materials, SiC can be thermally oxidized to form native $SiO_2$ dielectric layers, the insulator for metal-oxide–semiconductor (MOS) applications. Extensive research has been carried out [1-6] to determine the oxidation mechanism of SiC, leading to a modified Deal-Grove model [2] based on the established model for Si oxidation[3]. Several parameters determining the $SiO_2$ thickness have been investigated and established, such as oxidation time [4], temperature [5,6] and substrate orientation [7]. In this paper, we present results isolating the doping dependence of thermal oxidation rates on n-type 4H-SiC. Such studies have been carried out extensively for Si-oxidation [8], while there are none for SiC. An understanding of the doping dependence of SiC thermal oxidation is critical for the engineering of MOS dielectrics on SiC, which typically contain regions of very different doping concentrations e.g. >$10^{18}$cm$^{-3}$ for source/drain regions vs. ~$10^{16}$cm$^{-3}$ for channel regions.

Ref. [2] illustrates the five steps in the SiC thermal oxidation process , governed by the following overall chemical reaction:

$$SiC + 1.5O_2 \leftrightarrow SiO_2 + CO \qquad (1)$$

The five steps are

1. Transport of molecular oxygen gas to the oxide surface
2. In diffusion of $O_2$ through the oxide film
3. Reaction with SiC at the oxide/SiC interface
4. Out diffusion of product gases (e.g., CO) through the oxide film and
5. Removal of product gases away from the oxide surface

In this modified Deal-Grove model, the thickness, X, of a grown oxide film is given by

$$t + \tau = \frac{X^2}{B} + \frac{X}{B/A} \qquad (2)$$

Where t is the oxidation time and $\tau$ is a time that accounts for any native oxide. In a more general sense, $\tau$ is an important parameter because it also accounts for anomalies in the early stage of oxidation, and provides a fitting factor to adequately describe these anomalies [9]. If the native oxide and the grown oxide thickness are comparable in value, neglecting $\tau$ will cause erroneous results in the use of the Deal-Grove model for predicting the oxide thickness. B is the parabolic rate constant and B/A is the linear rate constant . From this equation, it can be seen that at small thicknesses, X, the linear term on the right-hand side dominates, whereas at large thicknesses, the quadratic, or parabolic term dominates.

At short oxidation times, there is almost no oxide present on the surface, so that there is no diffusion process occurring through the oxide. Thus, the linear term, B/A, is dependent on the kinetics of the interfacial reactions producing $SiO_2$ at the SiC surface (Step 3). Because oxidation takes place in a $O_2$ ambient, $C_{O2}^* >> C_{CO}^*$ , and the linear rate constant can be written as [2]

$$\frac{B}{A} \approx \frac{C_{O2}^*}{N_0} K_f \qquad (3)$$

where $N_0$ is the number of oxidant molecules incorporated into a unit volume of oxide layer (which is a constant) , $K_f$ is the rate constant of the forward reaction in equation (1) .

At longer oxidation times, the reactants have to get to and from the SiC/SiO$_2$ growth interface by diffusing through the thick oxide layer. Thus, the parabolic term, B, is determined by the diffusivity of the growth species in the grown SiO$_2$ films. The parabolic rate constant(B) is given by[2]

$$B = \frac{(K_f C_{O_2}^* - K_r C_{COX}^*)/N_0}{\frac{1.5 K_f}{D_{O_2}} + \frac{K_r}{D_{CO}}} \qquad (4a)$$

Where $C_{O_2}^*$ is the equilibrium concentration of the oxygen gas , which is a constant (corresponding to a partial pressure of 1 atm), $D_{O_2}$ is the diffusion coefficient in the oxide for O$_2$ and $D_{CO}$ is the diffusion coefficient in the oxide for CO. $K_r$ is the rate constant for the reverse reaction in equation (1) .

If oxygen in-diffusion (Step 2) is the rate controlling diffusion step , or $K_f/D_{O_2} \gg K_r/D_{CO}$ , then the parabolic rate constant is

$$B = \frac{C_{O_2}^*}{1.5 N_0} D_{O_2} \qquad (4b)$$

If CO out-diffusion is the rate controlling diffusion step (step 4), the parabolic rate constant is

$$B = \frac{K_f C_{O_2}^*}{K_r N_0} D_{CO} \qquad (4c)$$

This framework will be used to interpret the doping-dependent oxidation results presented below.

**Experimental:**

n-type(Nitrogen doped) 8mm×8mm 4H-SiC samples, cut 8$^0$ off axis from the [0001] plane were used. Samples were cleaned using trichloroethylene, acetone, methanol, and 5% HF solution in successive steps. Immediately thereafter, atomic force microscopy measurements were used to determine the R.M.S roughness in a 5µm×5µm area at several points on each sample, revealing uniform roughness (0.5 to 0.8nm) for all samples. A Keithley 590 C-V analyzer was used to measure the capacitance-voltage (C-V) characteristic of each sample and the corresponding doping density was extracted [10]. The data reported below are for samples cut from five wafers , henceforth referred to as wafers A, B, C, D and E, with decreasing doping concentrations, 9.3×10$^{18}$/cm$^3$, 7.4×10$^{18}$/cm$^3$, 2.1×10$^{17}$/cm$^3$, 1.6×10$^{17}$/cm$^3$ and 6.5×10$^{15}$/cm$^3$ respectively. Sample A and B are directly cut from the wafer whereas samples C,D and E have 20µm thick epilayers on them with the respective doping as indicated above. The low doped epitaxial samples(C, D and E) were grown by chemical vapor deposition (CVD) using conditions described elsewhere [11]. X-ray rocking curves show FWHM varying from 7-12 arc-sec to 30-40 arc-sec as the n-type 4H-SiC doping density varies from 6.5x10$^{15}$cm$^{-3}$ to $9.3 \times 10^{18} cm^{-3}$, indicating a degradation of crystal quality with increasing doping. The effect of this degradation in crystal quality with increasing doping is explained later in conjunction with increased oxidation rate.

The samples were then placed in 5% HF solution for 5 minutes to remove the native oxide and rinsed in de-ionized water. They were then loaded in an oxidation furnace for dry thermal oxidation with oxygen flow rate maintained at 9.3SLM. Oxidation was carried out at three different temperatures : $1000^0C$, $1100^0C$ and $1190^0C$. The temperature was not increased beyond $1200^0C$ since ionic oxidant transport and/or crystallization can occur on the oxide surface at this temperature [5]. At each temperature, oxidation was done for 2, 4, 6 and 8hours respectively. After each experiment, the samples were unloaded from the furnace at room temperature. The RMS roughness was measured using AFM in a 5μm×5μm area at several points on top of each sample, revealing oxide surface roughness 0.3-0.5nm. After all characterizations were performed, the oxide was stripped using HF. The interfacial RMS roughness was then measured to be ~0.5-0.8nm by AFM. These low roughness values allow us to ignore the influence of surface and interfacial roughness on our thickness measurements by ellipsometry [12].

The oxide thickness was measured using ellipsometry and mercury-probe C-V analysis in the accumulation region [10]. The accumulation capacitance, C, which is related to the oxide thickness, X, is given by

$$c = \frac{\varepsilon A}{X}$$

Where, $\varepsilon$ is the dielectric constant of $SiO_2$ (~3.7) and A is the mercury contact area. For ellipsometry, 9 points were measured on the sample to obtain the oxide thickness. The error bars here were calculated from the thickness difference of these points. For C-V, the precision of our capacitance meter is $\pm 0.1pF$, and is the major source of error for this measurement. Thicknesses obtained from these two techniques are consistent within the error bar limit (Figure 1). Oxide thickness extracted from C-V falls within the error bar limit of the thickness measured by ellipsometry, illustrating the self-consistency of the 2 measurement techniques.

**Results and Discussion:**

Figures 1a, 1b and 1c show the oxide thickness measured by ellipsometry, X, of the Si face as a function of oxidation time and doping for $1000^0C$, $1100^0C$ and $1190^0C$ respectively. From these plots, it is seen that the oxide thickness increases with time [4], temperature [5,6] and with doping concentration. Further, as all the samples are $8^0$ off -axis from the [0001] plane, we are confident that the variation in

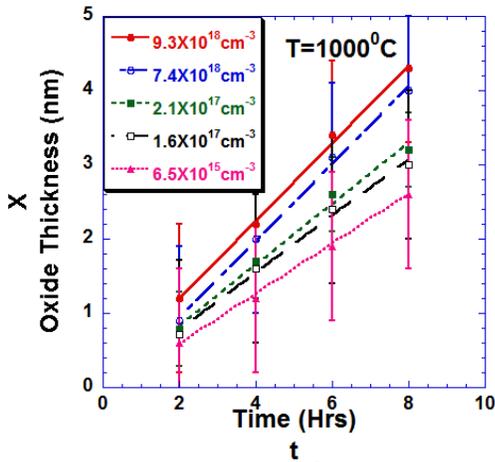

**Fig 1(a): Oxidation at 1000⁰C with 9.3SLM O₂ flow.**

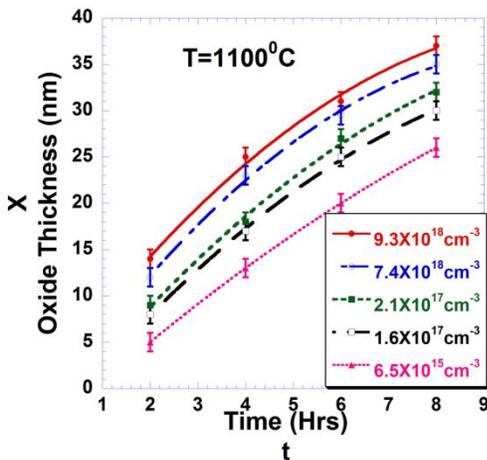

**Fig 1(b): Oxidation at 1100⁰C with 9.3SLM O₂ flow.**

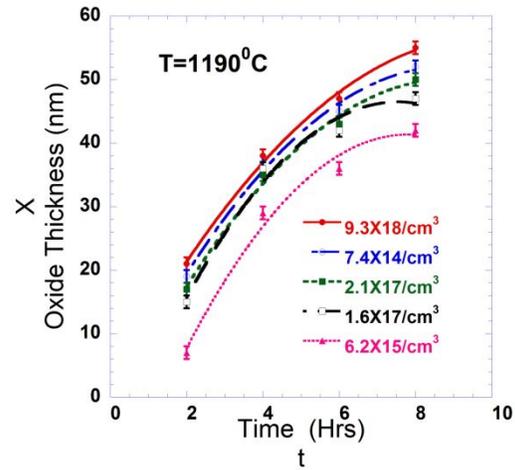

**Fig 1(c): Oxidation at 1190⁰C with 9.3SLM O₂ flow.**

oxide thickness with doping is not masked by variations in substrate orientation [7]. Another possible contribution to differences in oxidation rate is the starting surface roughness, similar for all samples in this study. This precludes surface roughness as a contributing factor to the observed variation in oxidation rate, leaving doping differences as the major factor accounting for the variations in oxidation rate.

The parabolic and linear rate constants are related to the activation energy by the following equations:

$$B = B_0 \exp(-E_B/KT) \qquad (5a)$$

$$B/A = (B/A)_0 \exp(-E_{(B/A)}/KT) \qquad (5b)$$

where $E_B$ and $E_{B/A}$ are the parabolic and linear activation energies, $B_0$ and $(B/A)_0$ are the parabolic and linear pre-exponential constants respectively and T is the absolute temperature. Figure 2 illustrates how these parameters were extracted, while accurately accounting for $\tau$, and is described in greater detail below. This procedure is shown for a single time

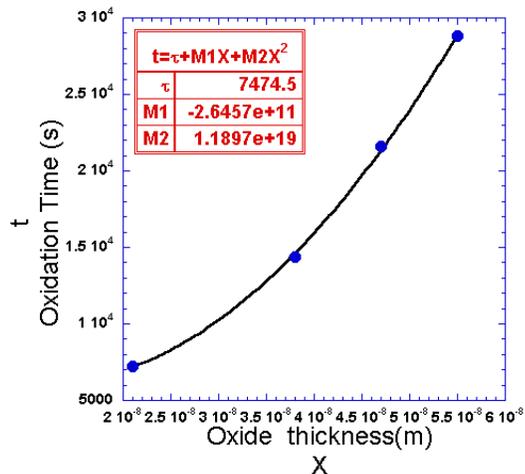

**Fig2(a):** 2$^{nd}$ order polynomial fit to extract τ at 1100$^0$C for the sample of doping density 1.6×10$^{17}$/cm$^3$ .

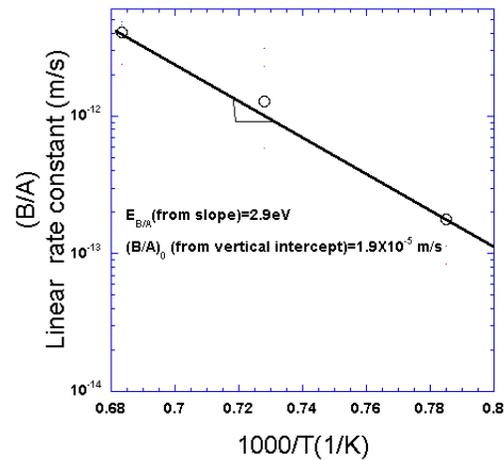

**Fig2(c):** Linear rate constant as a function of Temperature for the sample of doping density 1.6×10$^{17}$/cm$^3$ .

First, the oxidation time (t) was plotted against X (Figure 2a) for a given time/temperature/doping series and was fit with a second order polynomial (equation 2). The constant in this fit corresponds to τ. Using this τ, we plot X against (t+τ)/X for each temperature and a linear fit was performed (Figure 2b), as described in [9] for the extraction of the linear and parabolic rate constants. The slope of this line gives B, while the vertical intercept gives A, from which B/A is deduced[9].

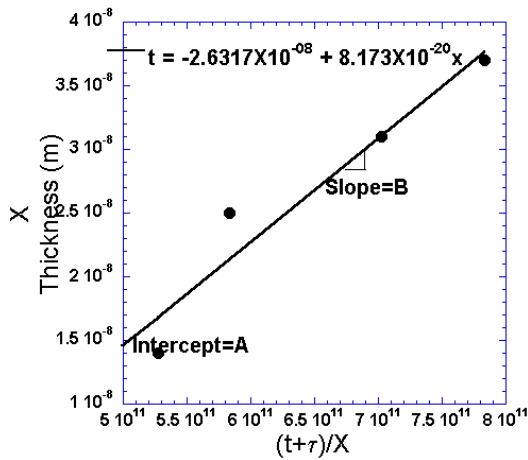

**Fig-2(b):** Linear fit to extract linear and parabolic rate constant at 1100$^0$C for the sample of doping density 1.6×10$^{17}$/cm$^3$ .

dependent series, although this was used for all the data in this paper.

B/A is then plotted as an Arrhenius function of T in figure 2(c), showing an increase of B/A with T, in agreement with ref [13]. From this Arrhenius plot, (B/A)$_0$ and E$_{B/A}$ are extracted (equation 5b). A similar procedure was used to extract B$_0$ and E$_B$ using equation 5a.

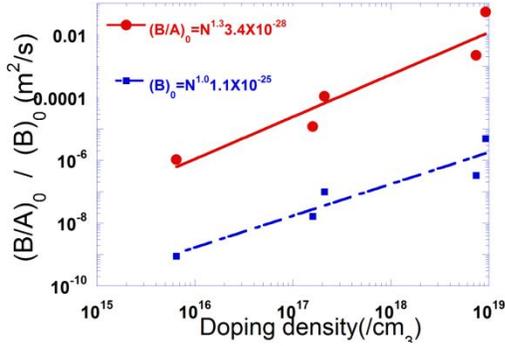

**Fig-3: Linear and parabolic rate constant as a function of doping concentration**

In this investigation, it was found that the oxidation rate is higher for highly doped samples compared to low doped samples. Higher oxidation rates are due to higher oxidation rate constants. In our dry thermal oxidation experiments we show an increase of both linear and parabolic rate constants with doping concentration. We will address each of these constants in greater detail in the discussion following.

**Rate Constant Pre-exponential Factors**

Figure 3 shows $(B/A)_0$ increasing from $\sim 10^{-6}$ to $\sim 10^{-2}$, as the doping concentration increases from $6.5 \times 10^{15}/\text{cm}^3$ to $9.3 \times 10^{18}/\text{cm}^3$. In equation (3), both $(C_{O2}^*)$ and $(N_0)$ are constant, as explained above, and cannot be affected by the increase of doping concentration. Thus, the only parameter that can cause an increase of the linear rate constant is $K_f$, which determines the interfacial reaction rate.

In n-SiC, nitrogen (N) doping leads to N-atoms substituting carbon (C) atoms in the

SiC lattice [14]. The smaller radius of the N-atom compared to the C atom results in a local mismatch in the SiC lattice, leading to the formation of defects in the SiC and degrading the crystal quality at high doping concentrations [15]. This is supported by our X-ray rocking curve data indicating an increase in FWHM from 7-12 arc-sec to 30-40 arc-sec as the doping concentration increases from $6.5 \times 10^{15}/\text{cm}^3$ to $9.3 \times 10^{18}/\text{cm}^3$. In other words, defect concentrations, originating from doping-induced lattice strain, increase with doping concentration [15]. These defect sites are more susceptible to oxidation than non-defect sites owing to a greater number of dangling bonds, and would lead to an increase in oxidation rate, and consequently, $K_f$, for thin oxide layers. In this study, the doping concentration is changed from $6.5 \times 10^{15}/\text{cm}^3$ to $9.3 \times 10^{18}/\text{cm}^3$ causing the surface density of the doping induced defects to increase by $(9.3 \times 10^{18}/6.5 \times 10^{15})^{2/3} \cong 100$, suggesting that the interfacial oxidation rate should increase by at least 100 times. $(B/A)_0$ increases from $10^{-6}$ to $10^{-2}$, or by $\sim 10^4$. This suggests that the increase in linear oxidation rate constant is due, at least in part, to the defects caused by the doping-induced lattice mismatch, and that the thermal oxidation of SiC is mediated by defects, in agreement with recent reports [16]. We derive the following empirical

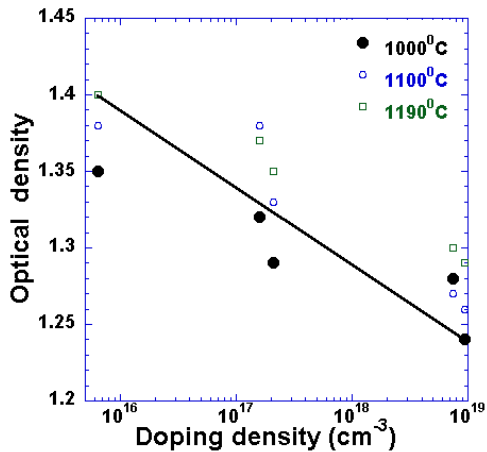

**Fig-4: Optical density of oxide as a function of doping density for different oxidation temperature.**

power-law relationship for $(B/A)_0$ by fitting to the data in Fig. 3:

$$(B/A)_0 = N^{1.3} 3.4 \times 10^{-28} \qquad (6)$$

As $(B/A)_0$ is proportional to the reaction rate constant $K_f$, equation (3) illustrates the change in interface reactivity with doping concentration. This power law dependence suggests that this reaction may be second order, as $(B/A)_0$ increases approximately as $(N^{2/3})^2$, although this point bears further investigation.

$B_0$, the parabolic pre-exponential factor, increases from $\sim 10^{-9}$-$10^{-6}$ m$^2$/s or by 1000 (Figure 3), as the doping increases from $6.5 \times 10^{15}$/cm$^3$ to $9.3 \times 10^{18}$/, by a factor $\cong 1400$, demonstrating a clear increase in the parabolic rate constant as a function of doping. The parabolic rate constant is dependent on the diffusion of gases in and out of the grown oxide layer (equations 4b and 4c), as discussed above. For this rate constant, if oxygen in-diffusion is the rate controlling step (equation 4b and Step 2), $C^*_{O_2}$ and $N_0$ are constant, and cannot be affected by increasing doping concentration. The only parameter that can increase the parabolic rate constant is $D_{O_2}$. Similarly, if CO out-diffusion is the rate controlling step (equation 4c and Step 4), $C^*_{O_2}$ and $N_0$ are again constant, while $K_f$ and $K_r$ will increase at the same rate, as SiC oxidation takes place under equilibrium conditions (shown in equation 1). Thus, the ratio $K_f/K_r$ will not change (equation 4c). In this case, the parabolic rate constant can only be increased by an increase of $D_{cO}$. In other words, regardless of the rate controlling diffusion step, the diffusivity of gases in the oxide must increase to account for the enhanced parabolic rate constants.

The diffusion coefficients of gases in the oxide (both O$_2$ and CO) can increase only if the oxide quality degrades, providing for a more porous oxide film, with voids to facilitate diffusion. Conversely, higher quality oxide should have lower diffusivity. Refractive index, or optical density, is an indication of the oxide density and quality. In our measurements of oxide thickness by ellipsometry, we found that the refractive index decreased from close to its ideal value, 1.4, for samples with the lowest doping concentrations (6.5×10^15/cm^3), to 1.24 for samples with the highest doping concentrations (9.3×10^18/cm^3) as shown in Figure 4. This indicates that increasing SiC doping concentration leads to a degradation in the oxide quality. The degradation in oxide density increases the diffusion coefficient in the oxide (both $D_{O_2}$ & $D_{co}$), ultimately accounting for the increase in $B_0$ with nitrogen doping in SiC. We derive the following empirical power law relationship for $B_0$ by fitting to the data in Fig. 3,

$$B_0 = N^{1.0} 1.1 \times 10^{-25} \qquad (7)$$

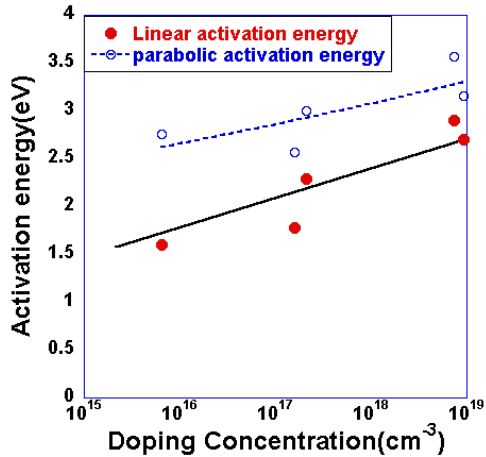

**Fig-5: Linear and parabolic activation energy as a function of doping density**

The linear change of $B_0$ with N indicates that our identification of doping-induced defects as the centers responsible for the enhanced oxidation rate in SiC is accurate. This is because $(B/A)_0$ is proportional to the oxide diffusivity, $D_0$ (equation 4c), which is expected to increase linearly with N.

## Rate Constant Activation Energies

Figure 5 shows the variation of linear ($E_{B/A}$) and parabolic ($E_B$) activation energies with doping concentration. $E_{B/A}$ varied from 1.6eV to 2.8eV, while $E_B$ from 2.7eV to 3.3eV, with both activation energies increasing with doping concentration. These values are consistent with those reported in the SiC literature [2],[17-18]. The doping dependence observed here may also explain the large spread in activation energies reported in the literature, from 0.6eV-3.6eV [2],[17-18]. This increase in activation energy is attributed to the difference in bond energy between the Si-C bond (2.82eV)[19] and the Si-N bond (4.26eV)[20]. For oxidation to take place in SiC, the Si-C bond must be broken in order for the new Si-O bond to be formed. Increasing nitrogen concentration increases the effective bond energy, as N-atoms occupy the C-atom positions [14]. Therefore, in order for oxidation to take place, an increased activation energy is required with increasing doping concentrations.

While this argument is clear for the increase of linear activation energy $E_{B/A}$, which deals with interfacial reaction of oxygen with SiC, where the bonds are broken, it is less obvious for parabolic activation energy $E_B$. This can be understood by noting that even after diffusing through the $SiO_2$ layer, the oxygen must react at the interface (Step 3) in order to generate an oxide layer whose thickness can be measured, even for thicker oxide films in the parabolic oxidation regime. For the interfacial reaction to occur, the Si-C and Si-N bonds must again be broken, presenting an energetic barrier to oxidation as in the linear oxidation regime. Despite this energetic barrier, the doping-dependence of $E_B$ is much less pronounced (2.7eV-3.3eV, or ~20% increase) than for $E_{B/A}$ (1.6-2.8eV, or ~75% increase), as diffusion kinetics undoubtedly influence the overall parabolic activation energy $E_B$. As the quality of the oxide degrades with doping concentration (discussed above), the energy barrier to gas diffusion will decrease (Steps 2 and 4), mitigating the influence of the increasing barrier presented by the Si-N bonds at the interface.

**Conclusion:**

In this study of 4H-SiC nitrogen-doping ($6.5 \times 10^{15}$/cm$^3$ to $9.3 \times 10^{18}$/cm$^3$) dependence of dry thermal oxidation, the oxidation rate was found to increase with doping concentration. This was accompanied by degradation in the oxide density as well.

These changes were attributed to doping-induced lattice mismatch, as well as the stronger Si-N bond as compared to the Si-C bond, and suggest that oxidation of SiC may, in some cases, be defect-mediated. The findings in this paper illustrate that doping variations play a crucial role in dry thermal 4H-SiC oxidation. This study has implications on the realization of uniform $SiO_2$ gate dielectrics on SiC MOS structures, which typically contain regions of very different doping concentrations.

**Acknowledgements**

This work was supported by the Southeast Center for Electrical Engineering Education (SCEEE), Grant No. SCEEE-09-001. The authors are grateful to the SCEEE Board of Directors for their interest and support of this research.

**Authors Biography:**

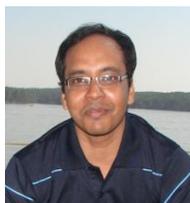

B.K.Daas, joined as a Graduate Research Assistant (PhD candidate) in University of South Carolina, USA in 2009. He received his Bachelors of Science  in Electrical Engineering fromBangladesh University of Engineering and Technology (BUET) ,Bangladesh, 2006.He worked as a lecturer in the department of EEE in Ahsanullah University of Science and Technology from Dec,2006 to Dec,2008 His research interest includes Wide band-gap semiconductor material, Device fabrication, Epitaxial Graphene Growth, Characterization and also graphene based electronic device fabrication

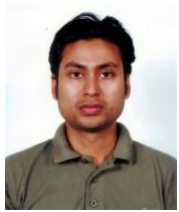

M.M.Islam, joined University of South Carolina as a Graduate Research Assistant (PhD candidate) in August'2008. He received his Bachelors of Science in Electrical Engineering from Bangladesh University of Engineering and Technology (BUET) in 2004. He worked as a switching network engineer in Telekom Malaysia from July'04 to August'08. Islam's research interests are in  Wide band-gap semiconductors device fabrication and characterization. His most recent work has focused on the optically controlled Silicon Carbide power electronic devices.

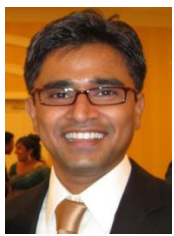

Iftekhar A Chowdhury joined Professor Sudarshan's research team in August 2005.  He completed his Bachelors of Science and Masters of Science in Applied Physics & Electronics at University of Dhaka. He is now a Graduate Research Assistant (PhD candidate) in Electrical Engineering at the Silicon Carbide Research Centre; working on the epitaxial growth of Silicon Carbide; for the power hungry next generation electronics

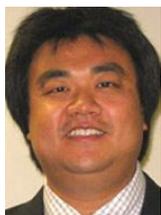 Feng Zhao (M'04) received the B.S. and M.Eng. degrees in materials science and engineering from Xi'an Jiaotong University, China, in 1997 and Nanyang Technological University, Singapore, in 2001, respectively, and the Ph.D. degree in electrical engineering from the University of Colorado at Boulder, CO, in 2004. From 2004 to 2008, he was with Microsemi Corporation as a Device Scientist to develop high frequency and high power SiC devices. In 2008, he joined the faculty of the University of South Carolina, where he is currently the Assistant Professor of electrical engineering. His research interests include SiC-based power electronics, photoconductive switches, and MEMS/NEMS sensors. He is a member of IEEE and MRS.

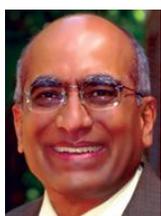 Dr.T.S.Sudarshan, Professor, Fellow IEEE, Department Chair, Carolina Distinguished Professor, University of South Carolina, USA. He received his B. Sc., (Physics, Mathematics, and Chemistry), University of Bangalore, Bangalore, India, May 1968, M. Sc., Physics (Solid state), University of Mysore, Mysore, India, June, 1970, M. A. Sc., Electrical Engineering (High Voltage Engineering) University of Waterloo, Ontario, Canada, August 1972, and Ph. D., Electrical Engineering (High Voltage Electrical Engr.), University of Waterloo, Ontario, October 1974.His research focous is on Photonics,Microelectronics and Nanoelectronics. His research interest includes Novel techniques of growth of silicon carbide (SiC) bulk and epitaxial films,SiC material and device processing – wafering, surface polishing, oxidation, mask technology, dopant diffusion, and metallization etc.

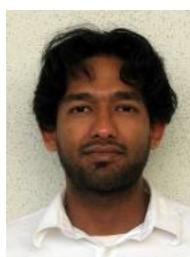 MVS Chandrashekhar "Chandra" obtained his B.S. from Worcester Polytechnic Institute in Worcester, MA in 2001, followed by a PhD from Cornell University, Ithaca, NY, Jan 2007, both in Electrical Engineering. He stayed on at Cornell till June 2009 as a post-doctoral associate, before joining the Electrical Engineering department at University of South Carolina, Columbia, SC as an Asssitant Professor. Born in India, raised in Singapore and educated in the United States, MVS Chandrashekhar has had broad exposure to the various types of energy problems that different economies face. This has focused his research towards novel, low cost, fast energy payback solutions to sustainable energy production, including dye-sensitized solar cells from SiC & GaN powders, photolytic/ photoelectrochemical hydrogen production as well as nanoelectronic materials such as graphene and SiC for electrical power management, sensing and emissions monitoring. His other interests lie in the physics of novel materials such as epitaxial graphene and also graphene based new electronic devices.